\documentclass{article}

 \usepackage[main, final]{neurips_2026_my}

\usepackage[utf8]{inputenc} 
\usepackage[T1]{fontenc}    
\usepackage{hyperref}       
\usepackage{url}            
\usepackage{booktabs}       
\usepackage{amsfonts}       
\usepackage{nicefrac}       
\usepackage{microtype}      
\usepackage{xcolor}         
\usepackage{graphicx}
\usepackage{caption}
\usepackage{multirow}
\usepackage{amsmath}
\usepackage{authblk}

\definecolor{cyanblue}{RGB}{40, 140, 200}
\hypersetup{
    colorlinks=true,
    citecolor=cyanblue,
    linkcolor=blue,
    urlcolor=black
}

\title{MindVoice: Reconstructing Intelligible Speech from Non-invasive Neural Signals with Pretrained Priors}

\author[ ]{Guangyin Bao}
\author[ ]{Taiping Zeng}
\author[ ]{Jianfeng Feng}
\author[ \textdagger]{Xiangyang Xue}
\affil[ ]{Fudan University}

\begin{document}

\maketitle

\begin{abstract}
Reconstructing continuous speech from non-invasive neural recordings is a fundamental problem for probing human auditory perception and building safe, scalable speech brain-computer interfaces. Despite recent progress, intelligible reconstruction remains elusive, as non-invasive recordings are inherently noisy, spatially blurred, and only partially preserve information about perceived speech. Existing methods directly map neural activity to entangled speech representations before synthesizing waveforms with neural vocoders, resulting in spectral-similar but unintelligible results.
To overcome these limitations, we introduce \textit{MindVoice}, a neuro-to-speech reconstruction framework that uses pretrained models to compensate for the incomplete semantic and acoustic information in neural recordings. \textit{MindVoice} disentangles reconstruction into two complementary pathways: one recovers high-level semantic content, while the other estimates fine-grained acoustic attributes. These inferred representations are then fused with powerful speech generation models and in-context voice cloning to synthesize natural and intelligible utterances.
Extensive experiments on EEG and MEG demonstrate that \textit{MindVoice} substantially outperforms existing methods on various metrics. These results show that pretrained priors provide a principled way to bridge the gap between noisy neural recordings and natural speech, highlighting a promising attempt for auditory neuroscience research and non-invasive speech brain-computer interfaces.
\end{abstract}

\section{Introduction}
Auditory perception lies at the core of human cognition. This multi-level neural process not only encompasses the semantic content of speech, but also conveys rich acoustic information such as pitch, intonation, and emotional prosody. For decades, researchers have sought to understand how the brain comprehends complex speech signals~\cite{hickok2007cortical, rauschecker2009maps, davis2003hierarchical}. Reconstructing original speech from neural activity provides direct evidence for uncovering the encoding mechanisms of auditory neural pathways, while also laying the groundwork for next-generation speech brain-computer interfaces. In the context of invasive settings, substantial progress has been made in speech-related decoding and reconstruction~\cite{mesgarani2014phonetic, wairagkar2025instantaneous, anumanchipalli2019speech, kunz2025inner, littlejohn2025streaming, metzger2023high}. Unfortunately, when using non-invasive neural signals with safety and broader applicability, such as electroencephalography (EEG) and magnetoencephalography (MEG), speech reconstruction remains a major technical challenge. This difficulty primarily stems from the low spatial resolution and massive noise of non-invasive recordings, which obscure auditory-evoked neural activity. At the same time, speech signals exhibit extremely fine-grained structures and spectrotemporal dynamics, making it difficult to establish precise and reliable mappings from scarce neural data that contain only weak and incomplete auditory information.

\begin{figure}[!t]
    \centering
    \includegraphics[width=1.0\linewidth]{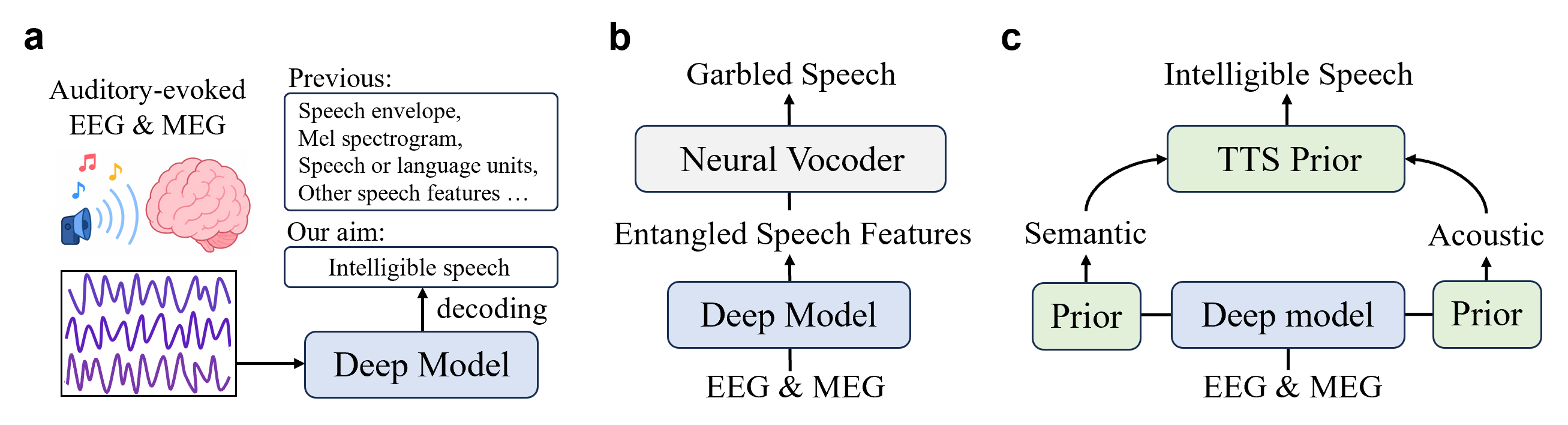}
    \captionsetup{font=small}
    \caption{\textbf{Decoding of auditory-evoked non-invasive neural recodings}. \textbf{a}, most extensive researches focus on decoding paraspeech information, whereas our goal is to directly reconstruct continuous and intelligible speech waveforms. \textbf{b}, existing EEG-to-speech methods align EEG with entangled speech representations and reconstruct speech using neural vocoders. \textbf{c}, our framework adopts a dual-stream architecture comprising semantic- and acoustic-level, which incorporates priors from pretrained models.}
    \vspace{-4mm}
    \label{intro}
\end{figure}

Extensive prior work has investigated non-invasive speech decoding from auditory-evoked neural activity. As illustrated in Figure~\ref{intro}{\color{blue} a}, these studies span a wide range of tasks, including speech envelope reconstruction~\cite{lalor2010neural, destoky2019comparing}, auditory-attention decoding~\cite{o2015attentional, xu2022decoding, accou2023decoding}, voice activity detection~\cite{sharon2020sound, ozdogan2025libribrain, landau20252025, de2025megconformer, xu2026shine}, prediction or retrieval of acoustic features such as mel-spectrograms~\cite{defossez2023decoding, li2024cross}, phonemes classification~\cite{de2025megconformer, xu2026shine}, and classification of words from small closed vocabularies~\cite{d2025towards, mcmurray2022decoding}. Collectively, these efforts establish important foundations and demonstrate the feasibility of decoding speech-related information from EEG or MEG signals. However, most existing studies focus on decoding paraspeech information, rather than directly reconstructing speech waveforms. As a result, they may provide limited insight into how the brain represents the acoustic and linguistic structure of natural speech. More recently, several pioneering studies~\cite{lee2024toward, lee2025enhancing} have attempted to reconstruct continuous speech from EEG signals. As shown in Figure~\ref{intro}{\color{blue} b}, these methods align neural activity with deep speech features and synthesize speech using neural vocoders. Nevertheless, the direct mappings learned from noisy neural recordings to entangled speech representations remain imprecise and therefore often yield reconstructed speech with limited intelligibility and natural fluency.

Non-invasive neural recordings suffer from low signal-to-noise ratios and limited spatial resolution, making continuous speech reconstruction a fundamentally underdetermined problem. Models are required to recover information-dense, structurally complex, and temporally sensitive speech from sparse, noisy, and unstable neural observations. As a result, relying solely on neural recordings is insufficient for achieving this target. Meanwhile, recent advances in generative AI have shown that large-scale pretrained models for speech, text, and multimodal data encode rich acoustic and linguistic priors, enabling strong performance in automatic speech recognition (ASR)~\cite{radford2023robust, barrault2023seamlessm4t, barrault2023seamless, gao2023funasr, shi2026qwen3}, text-to-speech generation (TTS)~\cite{du2024cosyvoice, chen2025f5, liao2024fish, hu2026qwen3}, and language modeling~\cite{yang2025qwen3, guo2025deepseek, grattafiori2024llama}. Inspired by recent progress in reconstructing images and videos from neural activity~\cite{benchetritbrain, chen2023seeing, scotti2023reconstructing, scotti2024mindeye2, li2024visual, chen2023cinematic, gong2024neuroclips, liu2024eeg2video, lu2025animate, yeung2025reanimating}, we suggest that continuous speech reconstruction should maximally leverage the priors encoded in such pretrained models to recover missing or semantically ambiguous speech information, thereby assembling intelligible speech from incomplete neural evidence.

In light of the above discussion, we propose \textit{\textbf{MindVoice}}, a neuro-to-speech reconstruction framework. Inspired by classical dual-stream accounts of speech perception~\cite{hickok2007cortical} and hierarchical auditory processing~\cite{davis2003hierarchical}, our framework adopts a dual-stream architecture comprising semantic-level and acoustic-level streams, as illustrated in Figure~\ref{intro}{\color{blue} c}. Rather than directly reconstructing speech from neural activity, this design decomposes the original problem into two lower-dimensional subproblems along complementary information axes, yielding more tractable alignment. Within each stream, we leverage priors encoded in pretrained models to compensate for incomplete speech information in neural signals, and further integrate the recovered semantic content with acoustic characteristics through text-to-speech generative models and in-context voice cloning to produce the final reconstructed speech.

We conduct extensive evaluations on widely used EEG~\cite{brennan2019hierarchical} and MEG~\cite{gwilliams2023introducing} datasets, showing that \textit{MindVoice} substantially outperforms prior neuro-to-speech reconstruction baselines across semantic accuracy, timbre, and speech quality metrics. Results across multiple datasets and evaluation splits demonstrate its generalizability. Qualitative comparisons and visualizations further corroborate these quantitative gains, while additional analyses and ablation provide insight into the model's interpretability and reconstruction preference. Together, these results demonstrate the effectiveness of our dual-stream design and the use of pretrained priors to compensate for missing speech information in neural recordings. They further establish \textit{MindVoice} as a state-of-the-art framework for reconstructing speech from EEG or MEG recordings, marking an important step in advancing neuro-to-speech reconstruction from garbled acoustic approximations toward intelligible speech.

\section{Related Works}

\textbf{Neural Speech Decoding}. 
Neural speech decoding encompasses a wide range of studies. Neural recordings can be categorized by acquisition method as invasive, such as ECoG and MEA, or non-invasive, such as EEG, MEG, and fMRI. They can also be grouped by the task paradigm used to elicit neural activity, including speech listening, imagined speech, and overt speech.
Invasive recordings provide high signal-to-noise ratios and fine spatiotemporal resolution, enabling precise capture of speech cortical dynamics. Accordingly, prior studies achieve high-accuracy and real-time decoding of listened and overt speech with invasive recordings~\cite{mesgarani2014phonetic, wairagkar2025instantaneous, anumanchipalli2019speech, littlejohn2025streaming, metzger2023high, willett2023high}, and begin to reconstruct imagined inner speech~\cite{kunz2025inner, jude2026decoding}.
In contrast, non-invasive recordings are safer, easier to collect, and more scalable, but are limited by lower signal quality and coarser spatial or temporal resolution. As a result, non-invasive speech decoding has more commonly focused on auxiliary speech-related information, such as acoustic features~\cite{lalor2010neural, destoky2019comparing, defossez2023decoding, li2024cross}, speech or semantic units~\cite{lee2023towards, de2025megconformer, xu2026shine, d2025towards, mcmurray2022decoding, abdulghani2023imagined, sato2024scaling}, and continuous speech features or language~\cite{lee2024toward, lee2025enhancing, yang2024mad, li2025brainecho, yang2026neuspeech, rastogi2025towards}, rather than direct speech waveform reconstruction.
Among non-invasive modalities, EEG and MEG are better suited for speech reconstruction than fMRI because their temporal resolution better matches speech dynamics. Across task paradigms, imagined speech faces the inherent challenge of temporal alignment with neural recordings; overt speech is confounded by articulation and movement artifacts, and prior studies~\cite{jo2024eeg} have shown that learning representations from such recordings often fails to outperform noise-level baselines.
Therefore, we focus on reconstructing listened speech from EEG and MEG.

\textbf{Speech Large Language Models}.
Speech LLMs have gained increasing attention in research communities. For automatic speech recognition~\cite{radford2023robust, barrault2023seamlessm4t, barrault2023seamless, gao2023funasr, shi2026qwen3, amodei2016deep, gulati2020conformer, baevski2020wav2vec}, recent studies leverage large-scale pre-training to improve transcription robustness across diverse acoustic conditions. For text-to-speech synthesis~\cite{du2024cosyvoice, chen2025f5, liao2024fish, hu2026qwen3, arik2017deep, ren2019fastspeech, renfastspeech, kim2020glow, kim2021conditional}, LLM-based speech generation has advanced expressive audio modeling by predicting discrete or continuous speech representations. To achieve personalized speech, many works~\cite{du2024cosyvoice, chen2025f5, liao2024fish, hu2026qwen3} further explore speech cloning techniques that adapt synthetic speech to target speakers using limited reference audio, either by conditioning generation on speaker embeddings or by modeling codec-based speech tokens. These methods improve speaker similarity, prosody similarity, and expressive controllability. In our work, we leverage the priors of speech LLMs to supplement the incomplete information in neural recordings.

\section{Methods}

\subsection{Overview}

\begin{figure}[!t]
    \centering
    \includegraphics[width=0.99\linewidth]{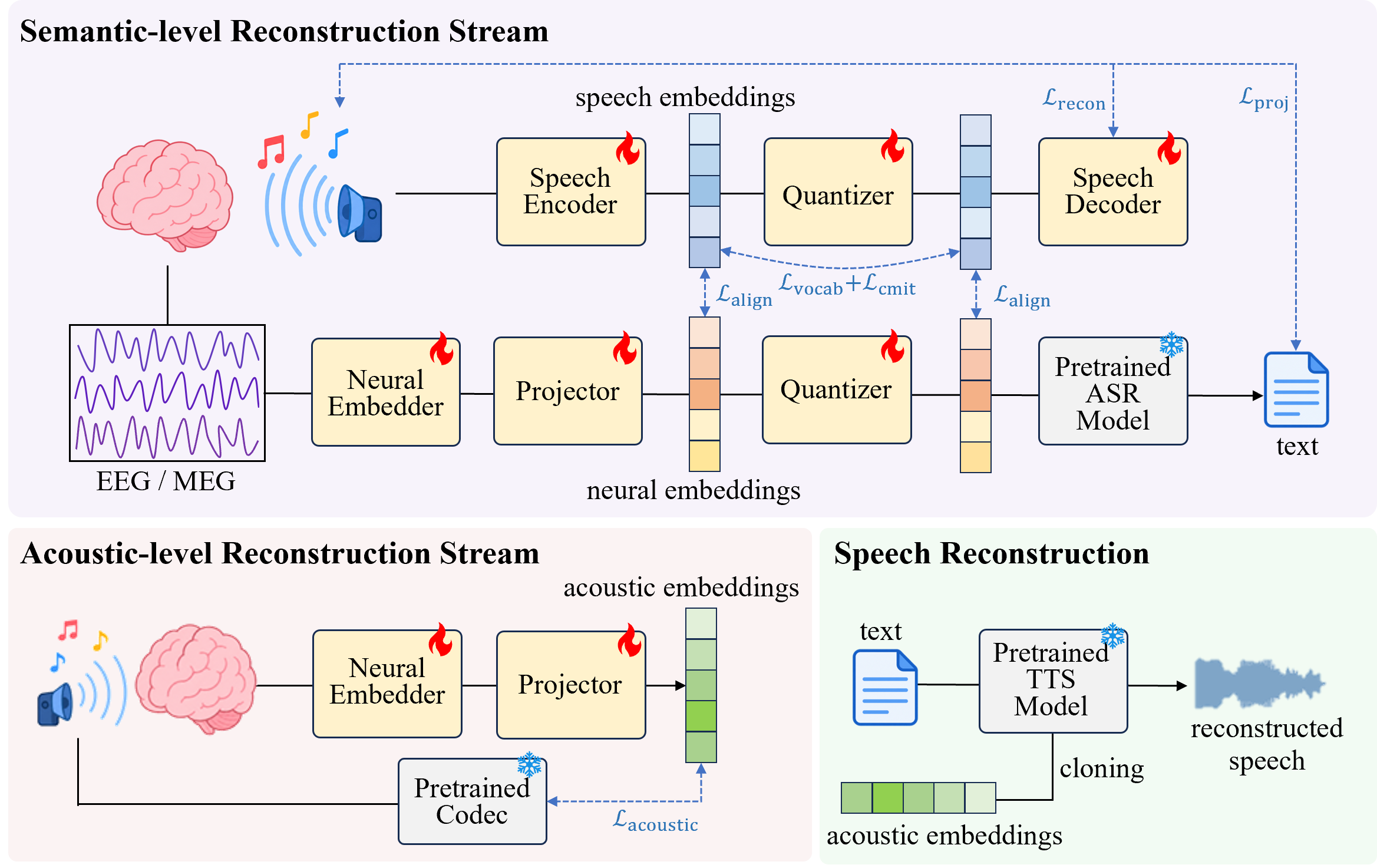}
    \vspace{-1mm}
    \caption{\textbf{Overview of the proposed MindVoice framework}.}
    \label{model}
\end{figure}

Speech is an intrinsically complex modality that carries heterogeneous information across multiple dimensions, including linguistic semantics, pitch and timbre, and fine-grained prosody. Consequently, directly regressing from neural recordings to speech representations in which these factors are entangled is challenging and often leads to limited reconstruction results.
To address this challenge, \textit{MindVoice} adopts a dual-stream reconstruction framework that predicts distinct dimensions of speech-related information, with each prediction process complemented by priors from pretrained generative models. As shown in Figure~\ref{model}, \textit{MindVoice} consists of three components: \textbf{1)~the Semantic-level Reconstruction Stream}, which leverages the language modeling prior of ASR models to infer and refine incomplete semantic information decoded from neural recordings; \textbf{2)~the Acoustic-level Reconstruction Stream}, which exploits acoustic priors from pretrained speech codecs, to obtain specific pitch and timbre cues implicitly encoded in neural recordings; and \textbf{3)~the Speech Reconstruction Branch}, which uses the semantic-level and acoustic-level results as constraints while injecting prosodic priors from pretrained TTS models into the final speech reconstruction process.

\subsection{Semantic-level Reconstruction Stream}
The accuracy of semantic content is crucial for speech, as it not only directly affects the intelligibility of the reconstructed results but may also influence key speech attributes such as prosody and temporal dynamics. Our Semantic-level Stream plays a key role in preserving this property.

For each paired EEG/MEG-speech segment $(\mathcal{X}_i, \mathcal{Y}_i)_{i=1}^{N}$, where $N$ denotes the number of training samples, we extract the transcript tokens $\mathcal{T}_i$ corresponding to the speech segment $\mathcal{Y}_i$ from the dataset annotations. Our semantic-level reconstruction stream consists of three modules: a neural signal embedder, a speech vector-quantized autoencoder, and a neuro-to-semantic aligner, which together generate the reconstructed text tokens $\hat{\mathcal{T}}$. 

\textbf{Neural Signal Embedder}. 
We treat EEG/MEG signals with temporal and channel dimensions as single-channel images, i.e., $\mathcal{X}_i \in \mathbb{R}^{1\times \mathrm{t}\times \mathrm{c}}$. 
Under this formulation, cascaded CNNs progressively compress the input and learn representations along the temporal and channel dimensions, respectively. 
A channel-wise MLP then projects the resulting image-like features into the neural latent space $\mathbb{R}^{\mathrm{t}'\mathrm{c}'\times d}$, where each $d$-dimensional neural token encodes activity within a local spatiotemporal region. 
Finally, we add learnable cosine positional encodings and apply a Transformer to model inter-token dependencies, producing the final neural embeddings $E_{\mathcal{X}_i}$.

\textbf{Speech Vector-Quantized Autoencoder}. 
We build the speech vector-quantized autoencoder to learn compact speech embeddings. 
Unlike raw speech waveforms, semantic information is not inherently continuous; therefore, we introduce a quantizer that maps continuous speech latents into discrete semantic tokens. 
These tokens constitute a speech vocabulary, enabling speech semantics to be modeled analogously to text.
Specifically, given a speech signal $\mathcal{Y}_i$, we first compute its mel-spectrogram $\mathcal{M}_i$. An encoder composed of convolutional layers, downsampling layers, and self-attention blocks then compresses $\mathcal{M}_i$ into latent embeddings $S_{\mathcal{Y}_i}$. 
To obtain discrete speech tokens, we construct a learnable small-capacity vocabulary and perform nearest-neighbor lookup between the continuous embeddings and the vocabulary vectors based on the $\ell_2$-distance. 
The selected vocabulary entries form the quantized latent embeddings $S_{\mathcal{Y}_i}^q$.
A decoder, roughly symmetric to the encoder, subsequently maps it back to the spectrogram space, yielding the reconstructed spectrogram, donated as $\hat{\mathcal{M}}_i$.
The training objective of the speech vector-quantized autoencoder can be defined as:
\begin{equation}
\begin{gathered}
\mathcal{L}_{\mathrm{vqae}}
= \mathcal{L}_{\mathrm{rec}} 
+ \alpha_1 \mathcal{L}_{\mathrm{vocab}}
+ \alpha_2 \mathcal{L}_{\mathrm{cmit}}, 
\\
\scalebox{0.93}{$
\mathcal{L}_{\mathrm{rec}} 
= \frac{1}{N} \sum\limits_{i=1}^N \bigl\| \mathcal{M}_i - \hat{\mathcal{M}}_i \bigr\|_2^2, 
~
\mathcal{L}_{\mathrm{vocab}} 
= \frac{1}{N} \sum\limits_{i=1}^N \bigl\| \mathrm{sg}[S_{\mathcal{Y}_i}] - S_{\mathcal{Y}_i}^q \bigr\|_2^2, 
~
\mathcal{L}_{\mathrm{cmit}} 
= \frac{1}{N} \sum\limits_{i=1}^N \bigl\| S_{\mathcal{Y}_i} - \mathrm{sg}[S_{\mathcal{Y}_i}^q] \bigr\|_2^2 .
$}
\end{gathered}
\end{equation}
Here, $\mathcal{L}_{\mathrm{rec}}$ encourages the reconstructed spectrogram $\hat{\mathcal{M}}_i$ to match the ground-truth spectrogram $\mathcal{M}_i$. 
The $\mathcal{L}_{\mathrm{vocab}}$ updates the vocabulary vectors by moving them toward the encoder latents, while the commitment loss $\mathcal{L}_{\mathrm{cmit}}$ encourages the encoder outputs to stay close to the selected vocabulary entries, preventing unstable switching across entries during training. 
$\mathrm{sg}[\cdot]$ denotes the stop-gradient operator, which is used with the straight-through estimator trick for training the quantizer.

\textbf{Neuro-to-Semantic Aligner}. 
We introduce an MLP-based projector to map the neural embeddings $E_{\mathcal{X}_i}$ into the pre-quantization speech embedding space, denoted as $\hat{S}_{\mathcal{Y}_i}$. 
The predicted continuous speech embeddings are then passed through the quantizer to obtain the predicted discrete speech embeddings $\hat{S}_{\mathcal{Y}_i^q}$. 
We jointly train the neural signal embedder and the projector, while fine-tuning the speech vector-quantized autoencoder, using the following objectives:
\begin{equation}
\mathcal{L}_{\mathrm{align}}
= \frac{1}{N} \sum_{i=1}^N \bigl\| S_{\mathcal{Y}_i} - \hat{S}_{\mathcal{Y}_i} \bigr\|_2^2
+ \frac{1}{N} \sum_{i=1}^N \bigl\| S_{\mathcal{Y}_i^q} - \hat{S}_{\mathcal{Y}_i^q} \bigr\|_2^2
+ \beta_1 \mathcal{L}_{\mathrm{vocab}}
+ \beta_2 \mathcal{L}_{\mathrm{cmit}}~.
\end{equation}
After training, the predicted discrete speech embeddings $\hat{S}_{\mathcal{Y}_i^q}$ can act as inputs of frozen ASR models for text transcription. Since the ASR models are pretrained with autoregressive text-token prediction on large-scale speech-text data, this neuro-to-text transcription process can leverage their language modeling capability as priors to complement incomplete semantic information within neural signals. To further encourage the alignment process to focus on semantic information, we also optimize the projector using a standard next-token prediction loss computed between the ASR model outputs and the ground-truth transcripts $\mathcal{T}_i$:
\begin{equation}
\mathcal{L}_{\mathrm{proj}} = - \sum_{i=1}^{N} \sum_{j=1}^{L_i} \frac{1}{NL_i}
\log \mathcal{P} \left( \mathcal{T}_j | \mathcal{T}_{<j}, \hat{S}_{\mathcal{Y}_i^q} \right).
\end{equation}
Here, $\mathcal{P}(\cdot)$ is the probability that ASR models assign to the correct next token given the input and previous tokens, while $\mathcal{T}_i = [\mathcal{T}_1, \mathcal{T}_2, ..., \mathcal{T}_{L_i}]$ denotes the $i$-th text with $L_i$ tokens in total. We do not fine-tune the ASR model, as the limited neural data appears insufficient to yield improvements. After staged training, the above modules form the semantic-level reconstruction stream.


\subsection{Acoustic-level Reconstruction Stream}

Beyond semantic content, speech signals encode rich paraspeech attributes such as pitch, timbre, emotion, and tone. Our acoustic-level reconstruction stream is therefore designed to extract pitch- and timbre-related embeddings that are critical for recovering these attributes.

To this end, we introduce a pretrained speech codec that follows the widely adopted codebook-based autoencoding paradigm in modern speech foundation models. Benefiting from large-scale pretraining, its codebook provides rich priors for encoding diverse paraspeech attributes at a fine-grained level. Given the speech segment $\mathcal{Y}_i$, the codec encodes it into latent embeddings $A_{\mathcal{Y}_i}$, which are then quantized to obtain $A_{\mathcal{Y}_i}^q$. In parallel, we train another neural signal embedder with the same architecture, together with another projector that differs only in its output dimensionality, to predict embeddings $\hat{A}_{\mathcal{Y}_i}$ from neural signals. Similarly, it is also quantized by the codec to yeild $\hat{A}_{\mathcal{Y}_i}^q$. We freeze the pretrained codec and optimize the embedder and projector under the following objective:
\begin{equation}
\mathcal{L}_{\mathrm{acoustic}}
= \frac{1}{N} \sum_{i=1}^N \bigl\| A_{\mathcal{Y}_i}^q - \hat{A}_{\mathcal{Y}_i}^q \bigr\|_2^2
- \gamma \frac{1}{N} \sum_{i=1}^N \log 
\frac{\exp\left( \mathrm{sim}(\hat{A}_{\mathcal{Y}_i},A_{\mathcal{Y}_i})/\tau \right)}
{\sum_{j=1}^N\exp\left( \mathrm{sim}(\hat{A}_{\mathcal{Y}_i},A_{\mathcal{Y}_j})/\tau \right)}
\end{equation}
Here, $\mathrm{sim}(\cdot)$ denotes cosine similarty, and $\tau$ is temperature factor for the contrastive loss. This process can be viewed as aligning neural signals with deep speech features. However, the resulting predictions often collapse toward averaged features, preserving only weak acoustic cues and little discernible semantic information. Quantization with pretrained codecs will map them to learned discrete units, thereby injecting priors and enhancing the discriminability of acoustic information.

\subsection{Speech Reconstruction}

Given the text $\hat{\mathcal{T}_i}$ reconstructed by the semantic-level stream and the acoustic embeddings $\hat{A}_{\mathcal{Y}_i}$ predicted by the acoustic-level stream, we leverage pretrained text-to-speech models for final speech reconstruction. 
Typically, a TTS model is composed of two parts: a Transformer $\mathcal{Q}(\cdot)$ that predicts speech latents from text tokens, and a decoder $\mathcal{D}(\cdot)$ that synthesizes waveforms from deep speech latent space. Based on this, the reconstruction process for final speech $\hat{\mathcal{Y}}_i$ can be formulated as:
\begin{equation}
q(\mathbf{z}_i \mid \hat{\mathcal{T}}_i, \hat{A}_{\mathcal{Y}_i})
=
\prod_{j=1}^{L_i}
\mathcal{Q}
\left(
z_{i,j}
\mid
z_{i,<j}, \hat{\mathcal{T}}_i, \hat{A}_{\mathcal{Y}_i}^q
\right),
\quad
\hat{\mathbf{z}}_i \sim q(\mathbf{z}_i \mid \hat{\mathcal{T}}_i, \hat{A}_{\mathcal{Y}_i}),
\quad
\hat{\mathcal{Y}}_i=\mathcal{D}\left( \hat{\mathbf{z}}_i \right).
\end{equation}
Here, we adopt in-context voice cloning, where the predicted acoustic embeddings $\hat{A}_{\mathcal{Y}_i}^q$ serve as acoustic prompts that guide the autoregressive generation of deep speech latents $\hat{\mathbf{z}}_i$. During this process, the TTS model leverages its learned priors to infer prosodic and temporal structures from the text, while using the acoustic embeddings as references for timbre- and pitch-related characteristics. Similarly, we do not fine-tune the TTS model using the discrepancy between the reconstructed speech and ground-truth, as such fine-tuning merely trades intelligibility for spectrogram similarity.

\section{Experiment Settings}

\textbf{Datasets and Preprocessing}. 
Following prior work on non-invasive speech decoding~\cite{defossez2023decoding}, we conduct experiments on the Brennan EEG dataset~\cite{brennan2019hierarchical} and the Gwilliams MEG dataset~\cite{gwilliams2023introducing}. Both datasets contain neural recordings collected while participants listened to continuous English narratives. The EEG dataset comprises 49 participants and 10.1 hours of recordings in total, whereas the MEG dataset comprises 27 participants and 49 hours of recordings. Our core preprocessing pipeline consists of notch filtering, band-pass filtering, and downsampling. The EEG/MEG recordings are then segmented at the sentence level and temporally aligned with the corresponding speech stimuli. For each EEG/MEG segment, we apply robust scaling, outlier clipping, and normalization. Given the pronounced non-stationarity of EEG and MEG signals, these segment-wise preprocessing steps are particularly important for reliable speech reconstruction. We perform two types of train-validation-test splits on the processed paired neural-speech segments using an 8:1:1 ratio: a random split and a strict sentence-based split. 
For more details, please refer to Appendix~\ref{appendix:datasets}.

\textbf{Baselines}. 
Since existing studies have rarely focused on directly reconstructing speech waveforms, we make every effort to identify suitable baselines for our method. A vanilla baseline is to regress speech features from neural recordings and then invert the predicted features into waveforms using a neural vocoder, which is commonly adopted in invasive speech reconstruction~\cite{wairagkar2025instantaneous}. To align this baseline with our method, we use mel-spectrograms as speech features and adopt the powerful BigVGAN-v2~\cite{leebigvgan} as the vocoder. Another suitable baseline is FESDE~\cite{lee2024toward}, which maps autoencoded EEG embeddings to VITS-style speech latents~\cite{kim2021conditional} through a normalizing-flow connector and then reconstructs speech using a self-trained HiFi-GAN decoder~\cite{kong2020hifi}.
More details can be found in Appendix~\ref{appendix:baselines}.

\textbf{Evaluation Metrics}. 
Our evaluation compares reconstructed speech against ground-truth speech. To assess similarity from multiple perspectives, we adopt a set of metrics covering multiple dimensions. At the spectral level, we use the mel-spectrograms MSE and mel-cepstral distortion (MCD)~\cite{kubichek1993mel} to measure spectral similarity. At the semantic level, we use HuBERT~\cite{hsu2021hubert} representation similarity to assess semantic consistency. We further transcribe the speech using Qwen3-ASR~\cite{shi2026qwen3}, which achieves human-level speech understanding, and calculate baseline-scaled BERTScore-F1~\cite{zhangbertscore} based on transcription results to quantify semantic accuracy. For timbre similarity, we compute representation similarity using WavLM~\cite{chen2022wavlm} embeddings. Finally, the MOS predictor score~\cite{saeki2022utmos} is used to evaluate overall speech quality. 
More details on metrics are provided in Appendix~\ref{appendix:metrics}.

\textbf{Implementation Details}.
We use the Whisper-base~\cite{radford2023robust} as the ASR model in our semantic-level reconstruction stream. 
We adopt the Codec and TTS model of FishSpeech-s1-mini~\cite{liao2024fish} as the codec in our acoustic-level reconstruction stream and final speech reconstruction model, respectively. Our experiments are conducted using 8 NVIDIA Tesla V100 GPUs. For detailed implementation settings and training hyperparameters, please refer to our Appendix~\ref{appendix:implementation}.

\section{Results}

\begin{table}[!t]
\centering
\captionsetup{font=small}
\caption{Quantitative experimental results. We report the averaged values under three random seeds.}

\resizebox{0.98\textwidth}{!}{
\begin{tabular}{cclcccccc}

\toprule
\multirow{2}{*}{Data} & \multirow{2}{*}{Split} & \multirow{2}{*}{Method} 
& \multicolumn{2}{c}{Spectral} & \multicolumn{2}{c}{Semantic} & Timbre
& Quality \\

\cmidrule(lr){4-5} \cmidrule(lr){6-7} \cmidrule(lr){8-8} \cmidrule(lr){9-9}

& & & mel-MSE~$\downarrow$ & MCD$~\downarrow$ & HuBert$~\uparrow$ & ASR-Bert$~\uparrow$ & WavLM$~\uparrow$ & MOS$~\uparrow$ \\
\specialrule{0.8pt}{0pt}{3pt}

\multirow{6}{*}{EEG}
& \multirow{3}{*}{Random}
& Vanilla       & 0.321 & 10.42 & 0.136 & 0.144 & 0.070 & 1.28 \\
& & FESDE       & \textbf{0.198} & \textbf{10.15} & 0.391 & 0.189 & 0.054 & 1.43 \\
& & MindVoice   & 0.413 & 10.50 & \textbf{0.752} & \textbf{0.379} & \textbf{0.664} & \textbf{4.26} \\
\cmidrule(lr){2-9}
& \multirow{3}{*}{Sentence}
& Vanilla       & 0.294 & 11.36 & 0.094 & 0.001 & 0.031 & 1.28 \\
& & FESDE       & \textbf{0.167} & \textbf{11.31} & 0.210 & 0.103 & 0.063 & 1.44 \\
& & MindVoice   & 0.501 & 11.38 & \textbf{0.447} & \textbf{0.120} & \textbf{0.245} & \textbf{3.36} \\

\specialrule{0.6pt}{0pt}{2pt}

\multirow{6}{*}{MEG}
& \multirow{3}{*}{Random}
& Vanilla       & 0.278 & 10.04 & 0.322 & 0.208 & 0.268 & 1.26 \\
& & FESDE       & \textbf{0.187} & 10.57 & 0.532 & 0.219 & 0.331 & 1.41 \\
& & MindVoice   & 0.445 & \textbf{10.01} & \textbf{0.829} & \textbf{0.441} & \textbf{0.777} & \textbf{4.35} \\
\cmidrule(lr){2-9}
& \multirow{3}{*}{Sentence}
& Vanilla       & \textbf{0.142} & \textbf{10.21} & 0.184 & 0.171 & 0.109 & 1.28 \\
& & FESDE       & 0.210 & 10.66 & 0.472 & 0.188 & 0.130 & 1.30 \\
& & MindVoice   & 0.457 & 10.35 & \textbf{0.820} & \textbf{0.324} & \textbf{0.758} & \textbf{4.34} \\

\bottomrule
\end{tabular}
}
\label{tab:results}
\end{table}

\subsection{Quantitative Experimental Results}

As shown in Table~\ref{tab:results}, we report the quantitative evaluation results of \textit{MindVoice} and the baseline methods across different datasets and split settings. Overall, \textit{MindVoice} achieves substantial improvements over the baselines on high-level metrics, including semantic similarity, timbre similarity, and perceptual speech quality. These results suggest that the speech reconstructed by \textit{MindVoice} better preserves semantic content, speaker-related timbre characteristics, and overall perceptual quality. In contrast, \textit{MindVoice} does not consistently outperform the baselines on spectral metrics. This is expected, since both baseline methods directly optimize the MSE of mel-spectrograms as the supervision objective for speech reconstruction, whereas \textit{MindVoice} does not optimize it during TTS-based generation. As a result, \textit{MindVoice} generally obtains worse mel-MSE than the baselines. Meanwhile, on MCD, which is closely related to mel-spectral distortion, \textit{MindVoice} performs comparably to the baselines. These observations suggest that \textit{MindVoice} prioritizes the reconstruction of high-level semantic and perceptual information rather than merely pursuing low-level spectral similarity. They further indicate a clear mismatch between spectral reconstruction errors and perceptual quality: lower spectral error does not necessarily translate into better speech intelligibility or perceptual naturalness. 

A further comparison between EEG- and MEG-based reconstruction reveals two consistent trends. First, both \textit{MindVoice} and the baseline methods achieve better reconstruction performance on MEG, which provides higher signal quality and a larger data scale. This finding is consistent with prior studies~\cite{defossez2023decoding}. Second, the data split has a substantial impact on reconstruction performance. Compared with the Random split, the more stringent Sentence split generally leads to an overall performance drop, especially on the smaller-scale EEG dataset. This drop is much smaller on MEG than on EEG, suggesting that higher-quality neural recordings and larger-scale training data can mitigate the generalization challenge posed by unseen sentences. Nevertheless, the advantage of \textit{MindVoice} remains pronounced under the Sentence split, indicating that the model does not merely memorize sentence-specific patterns but instead learns representations that generalize to unseen speech content. The comparable performance of \textit{MindVoice} on MEG-Random and MEG-Sentence further suggests that, given sufficient data quality and scale, our framework could learn a relatively stable neural-to-speech mapping beyond the specific sentences observed during training.

\subsection{Visualization and Qualitative Results}

\begin{figure}[!t]
    \centering
    \includegraphics[width=0.99\linewidth]{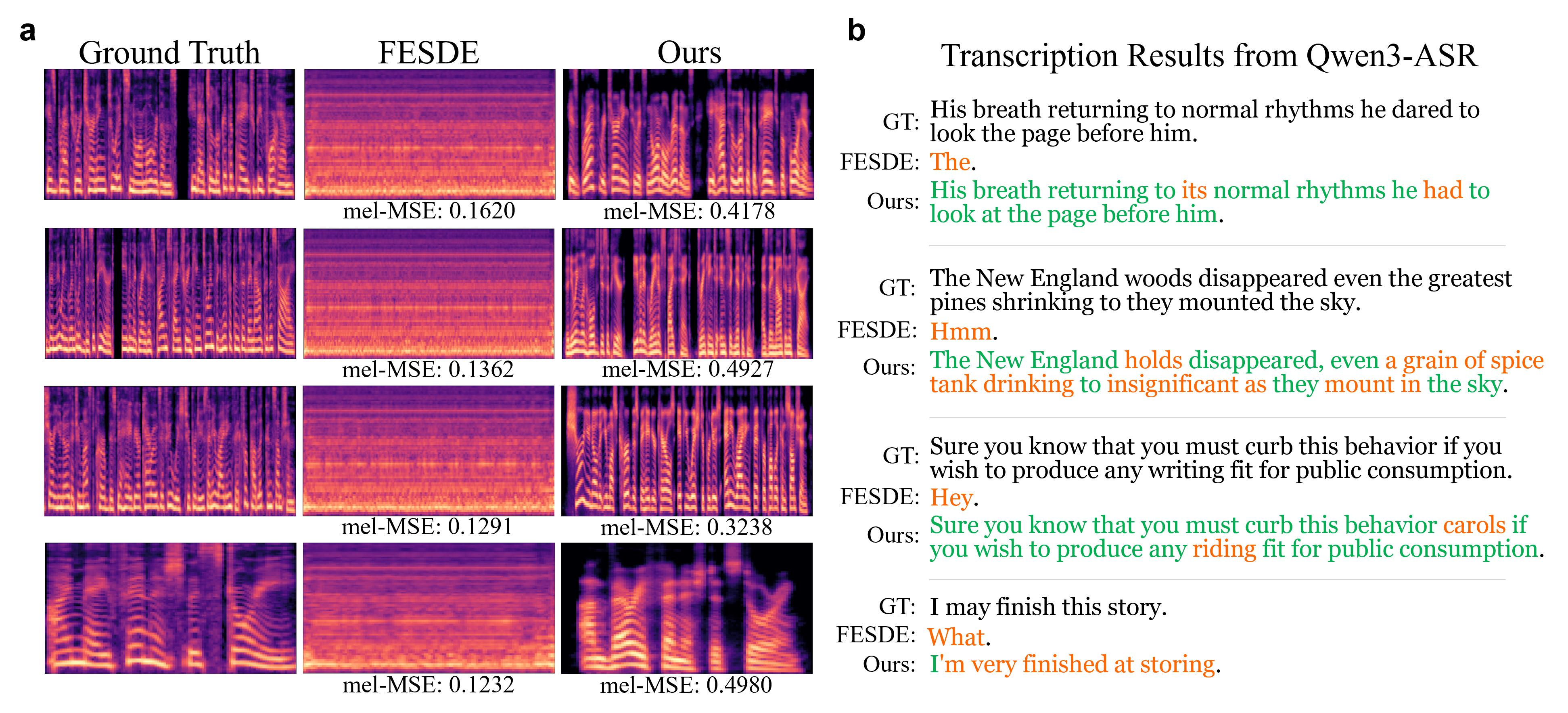}
    \vspace{-1mm}
    \captionsetup{font=small}
    \caption{\textbf{Mel spectrogram and transcription results}. \textbf{a}, comparison of mel spectrograms obtained by different methods with the ground-truth. \textbf{b}, transcription results of ground-truth speech, FESDE reconstructed speech, and our reconstructed speech using the Qwen3-ASR. More results can be found in Appendix~\ref{appendix:qualitative}.}
    \label{vis}
\end{figure}

We further provide qualitative evidence in Figure~\ref{vis}. Specifically, we visualize the mel-spectrograms of speech reconstructed by different methods and display the corresponding transcriptions obtained with Qwen3-ASR~\cite{shi2026qwen3}. As shown in Figure~\ref{vis}{\color{blue} a}, although the baseline FESDE achieves a substantially lower mel-MSE than \textit{MindVoice}, its reconstructed spectrograms are heavily over-smoothed. The energy distribution is overly uniform over time, and fine-grained vertical structures, harmonic patterns, transient changes, and local spectral details are largely absent. This suggests that existing methods primarily capture coarse spectral energy patterns, while failing to recover the dynamic acoustic structures that are critical for intelligible speech. In contrast, despite its higher mel-MSE, \textit{MindVoice} produces spectrograms that are visually more consistent with the ground truth, preserving periodic harmonic structures, temporal energy variations, and silent intervals, which are closely associated with linguistic content intelligibility and perceptual naturalness.

This observation is further supported by transcription results in Figure~\ref{vis}{\color{blue} b}. Speech reconstructed by \textit{MindVoice} contains recognizable linguistic content and can be understood by the human-level ASR model, whereas speech reconstructed by FESDE is mostly transcribed into short and semantically uninformative fragments. These results provide evidence for the mismatch observed in Table~\ref{tab:results}: point-wise spectral errors do not necessarily reflect perceptual quality or semantic intelligibility. 
Moreover, the errors made by \textit{MindVoice} are not random noise-like artifacts, but often take the form of semantically related approximations. As illustrated by the transcription examples, \textit{MindVoice} could preserve the main sentence structure, partial keywords, and grammatical patterns, although it may still introduce word substitutions, insertions, or semantic drift.

\subsection{Model Interpretability and Preference}

\begin{figure}[!t]
    \centering
    \includegraphics[width=0.99\linewidth,trim=1 2 1 1,clip]{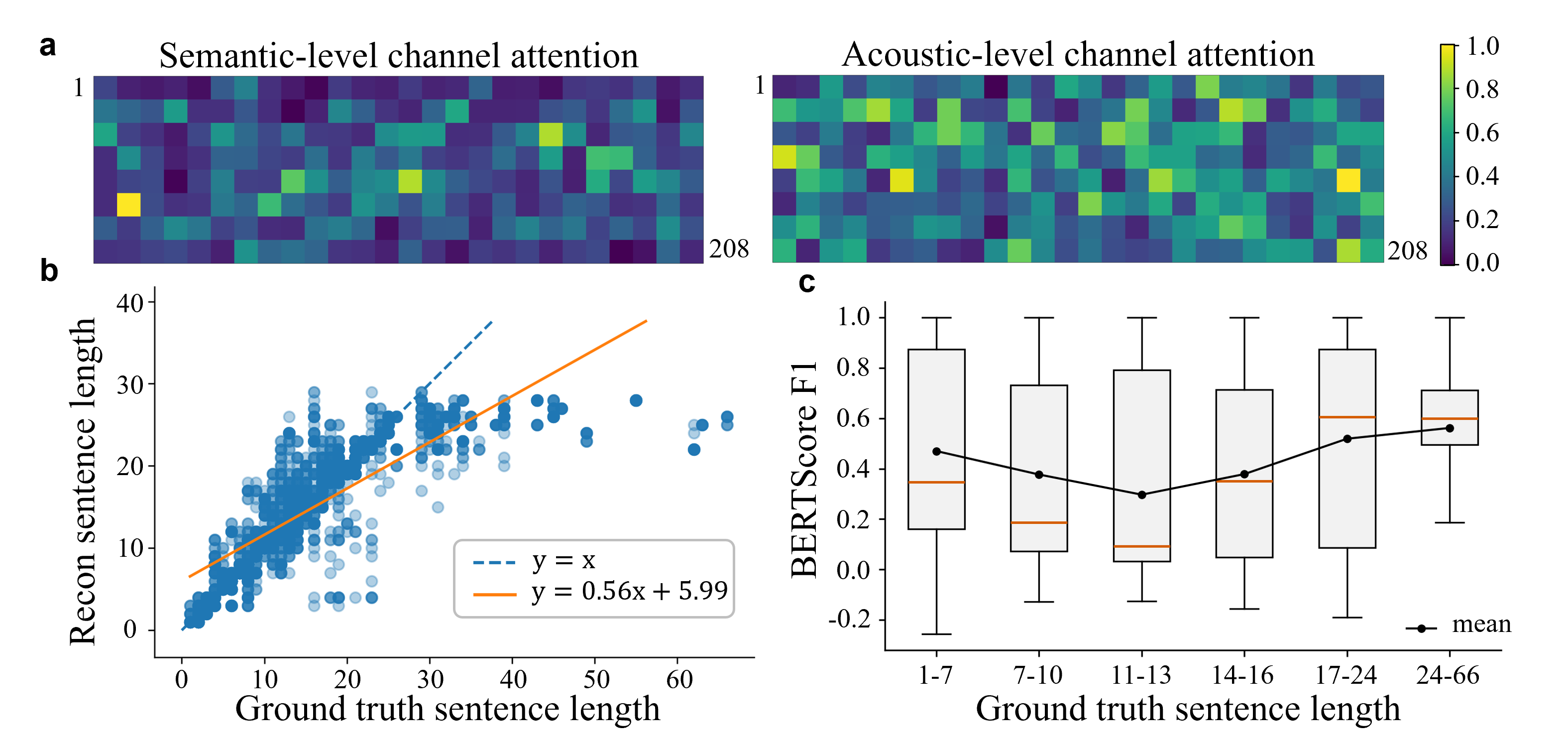}
    \vspace{-1mm}
    \captionsetup{font=small}
    \caption{\textbf{Analyses of model interpretability and preference}. \textbf{a}, channel heatmap for semantic- and acoustic-level stream. Each pixel represents a neural signal channel. \textbf{b}, sentence length regression analysis between reconstructions and ground truth. \textbf{c}, baseline-scaled BERTScore-F1 across sentence length groups.}
    \label{perfer}
\end{figure}

We analyze model interpretability in Figure~\ref{perfer}{\color{blue} a} by visualizing the channel-wise attention over the input MEG signals for the semantic- and acoustic-level streams. The semantic stream produces sparser high-response regions concentrated on a limited number of channels, whereas the acoustic stream shows a more distributed pattern with elevated responses across a broader set of channels. This indicates that semantic and acoustic reconstruction exploit partially distinct neural information, and that the model adaptively learns neural representations for different representation targets.

We further investigate the semantic reconstruction preferences of \textit{MindVoice}. As shown in Figure~\ref{perfer}{\color{blue} b}, the reconstructed sentence length is positively correlated with ground-truth length, indicating that the model captures length variations in semantic content. However, we still observe a convergence effect in its outputs. The fitted slope is 0.59, below the ideal identity line, suggesting a compression effect for most sentences, especially longer ones. This may arise because neural signals provide incomplete semantic information for equivalent-length reconstruction. For shorter samples, the model appears to rely more on pretrained priors to supplement missing information, leading to a large regression intercept. 
Figure~\ref{perfer}{\color{blue} c} shows the distribution of BERTScore-F1 across different sentence-length groups. Semantic reconstruction accuracy is above average for both short and long sentences. This pattern may be attributed to two factors: short sentences are generally easier to decode, whereas longer sentences may receive higher BERTScore values due to the presence of a few correctly reconstructed key words.
For additional analyses, please refer to our Appendix~\ref{appendix:perference}.

\subsection{Ablation Studies}

\begin{table}[!t]
\centering
\captionsetup{font=small}
\caption{Ablation results. We conduct experiments under the MEG dataset with the Random split.}

\resizebox{0.90\textwidth}{!}{
\begin{tabular}{lcccccc}
\toprule
\multirow{2}{*}{Method} 
& \multicolumn{2}{c}{Spectral} & \multicolumn{2}{c}{Semantic} & Timbre & Quality \\
\cmidrule(lr){2-3} \cmidrule(lr){4-5} \cmidrule(lr){6-6} \cmidrule(lr){7-7}
& mel-MSE~$\downarrow$ & MCD$~\downarrow$ & HuBert$~\uparrow$ & ASR-Bert$~\uparrow$ & WavLM$~\uparrow$ & MOS$~\uparrow$ \\
\toprule
MindVoice           & 0.445 & 10.01 & 0.829 & 0.441 & 0.777 & 4.35 \\
w/o Acoustic        & 0.458 & 10.13 & 0.807 & 0.441 & 0.732 & 4.35 \\
w/o Vocabulary      & 0.540 & 10.59 & 0.617 & 0.294 & 0.624 & 4.23 \\
w/o MEG             & 1.061 & 11.25 & 0.243 & 0.110 & 0.319 & 3.98 \\
\bottomrule
\end{tabular}
}
\label{tab:ablation}
\end{table}

We conduct ablation experiments to examine the contribution of each design in our \textit{MindVoice}. Specifically, we first remove the acoustic-level reconstruction stream and instead perform reconstruction directly from the semantic stream outputs. We then ablate the speech vocabulary in the semantic-level reconstruction stream, replacing it with a standard speech autoencoder. Moreover, suggested by previous work~\cite{jo2024eeg}, we replace the MEG signal with random Gaussian noise as the input of trained \textit{MindVoice}, which tests whether the reconstructed semantics are indeed derived from MEG signals rather than artifacts of an insufficiently rigorous evaluation protocol.
The ablation results are shown in Table~\ref{tab:ablation}. It can be seen that the semantic-level and acoustic-level streams in \textit{MindVoice} are only weakly coupled. After ablating the acoustic-level stream, only representation similarity metrics such as HuBert and WavLM show slight decreases, while ASR-BERTScore-F1, which more strictly evaluates semantic content reconstruction, remains unchanged. In contrast, ablating the speech vocabulary substantially degrades semantic-level reconstruction results, which further affects both spectral and timbre metrics. In addition, speech quality is mainly guaranteed by the pretrained TTS models in our framework, and therefore does not vary substantially as long as intelligible content can be reconstructed. Finally, in the absence of MEG input, the model degenerates into producing fixed short phrases, leading to significant drops across all metrics. This indicates that neural signals provide critical information for speech reconstruction.

\vspace{-1mm}
\section{Discussion}

Our \textit{MindVoice} aims to reconstruct continuous and intelligible speech from non-invasive neural signals. It partially succeeds in recovering participants’ auditory perceptual experiences and establishes a new state-of-the-art baseline for this task. Nevertheless, our study has several limitations. First, the reconstruction success rate remains limited. The best baseline-scaled BERTScore-F1 of 0.441 indicates that most reconstructions still contain substantial semantic discrepancies. This is an inevitable consequence of incorporating pretrained priors: when the neural signals do not provide sufficient information for precise decoding, the model may produce considerable generative hallucinations. Therefore, any model output should be interpreted with caution. Second, our reconstruction objective emphasizes overall semantic and timbre similarity, but does not guarantee the correctness of semantic or acoustic units at each time step. As a result, our \textit{MindVoice} may not be suitable for studies that require fine-grained temporal fidelity in these attributes. Finally, our \textit{MindVoice} has a clearly defined scope. It is currently applicable only to non-invasive neural signals evoked by auditory perception, and its performance on neural signals associated with overt speech or imagined speech remains unclear. Future work may therefore explore several directions, including improving reconstruction accuracy while reducing hallucinated content, achieving temporally aligned fine-grained speech reconstruction, and extending the proposed framework to overt-speech and imagined-speech neural signals toward a practical non-invasive speech brain-computer interface.

\vspace{-1mm}
\section{Conclusion}
\vspace{-1mm}

In this paper, we present \textit{MindVoice}, a framework for reconstructing continuous and intelligible speech from auditory-evoked non-invasive neural signals. By decoupling semantic and acoustic reconstruction and leveraging the priors of pretrained models, \textit{MindVoice} achieves substantial performance improvements. Evaluations on two datasets show that \textit{MindVoice} successfully reconstructs speech and establishes state-of-the-art performance across various metrics. Further experiments and analyses reveal the sources of performance gains and its reconstruction preferences. 
Overall, our study achieves reconstruction from garbled acoustic approximations toward intelligible speech.

\section*{Broader Impacts}
Our study reconstructs participants’ auditory experiences from neural signals. To the best of our knowledge, this work represents one of the earliest attempts to directly reconstruct speech waveforms from non-invasive neural recordings. By formulating the task of reconstructing intelligible continuous speech from non-invasive neural signals and presenting a reasonably effective baseline method, our work provides a useful starting point for future research in this direction. More broadly, this study offers potential tools and preliminary evidence for investigating neural mechanisms underlying auditory processing, as well as for developing future speech brain–computer interfaces.


{
\medskip
\small
\bibliography{ref}
\bibliographystyle{unsrt}
}

\newpage
\appendix

\section{Additional Experiment Settings}
\subsection{Datasets and Preprocessing}
\label{appendix:datasets}

\paragraph{Brennan EEG dataset.}
Brennan EEG dataset is a naturalistic listening EEG dataset introduced by Brennan and Hale~\citep{brennan2019hierarchical}, which was designed to study how hierarchical linguistic structure guides rapid predictions during spoken language comprehension. The dataset contains electroencephalography recordings collected while participants passively listened to the first chapter of \textit{Alice's Adventures in Wonderland}. The stimulus is a 12.4-minute English audiobook recording, divided into 12 speech segments, and each word is associated with timing information and linguistic variables used in the original analyses. The publicly released dataset comprises 49 human EEG recordings acquired at the University of Michigan Computational Neurolinguistics Lab using 61 active electrodes and a Brain Products actiCHamp amplifier, sampled at 500 Hz with an online band-pass range of 0.1--200 Hz. After listening, participants completed an 8-question comprehension questionnaire to verify attention to the story. Following the original study, a subset of participants was used for the main analysis after excluding recordings with low comprehension performance or high noise, making this dataset a compact but well-annotated benchmark for our research.

\paragraph{Gwilliams MEG-MASC dataset.}
Gwilliams MEG-MASC dataset is a large-scale magnetoencephalography dataset for evaluating neural responses to naturalistic speech~\citep{gwilliams2023introducing}. MEG-MASC consists of raw MEG recordings from 27 English-speaking adults who listened to approximately two hours of naturalistic English stories. The speech materials are four fictional stories selected from the Manually Annotated Sub-Corpus (MASC): \textit{LW1}, \textit{Cable Spool Boy}, \textit{Easy Money}, and \textit{The Black Willow}. In the standard design, each participant completed two matched one-hour sessions, with the same four stories repeated across sessions, enabling both signal averaging and matched train/test evaluation. MEG signals were recorded with a 208 axial-gradiometer KIT system at 1,000 Hz and online band-pass filtered between 0.01 and 200 Hz while participants listened through binaural tube earphones. To encourage attention, participants answered two-alternative comprehension questions roughly every three minutes, achieving high average accuracy in the original study. The dataset provides precise onset and offset timestamps for words and phonemes, is organized according to the Brain Imaging Data Structure (BIDS), and includes publicly available MEG, speech, text, metadata, and analysis code. These properties make MEG-MASC particularly suitable for our research.

\paragraph{EEG preprocessing and data splits.}
For the Brennan EEG dataset, we first constructed sentence-level EEG-speech pairs from the original continuous recordings. The speech files were concatenated in temporal order, resampled to 16 kHz, transcribed with WhisperX, and force-aligned to obtain sentence- and word-level timestamps. The resulting sentence boundaries were then used to segment the continuous EEG signals. For each subject, the raw EEG was loaded from the MATLAB files, notch-filtered at 60 Hz, band-pass filtered between 0.5 and 99 Hz, and resampled from 500 Hz to 200 Hz. We applied robust scaling channel-wise, clipped extreme values to the range $[-10, 10]$, normalized by the maximum absolute amplitude, and standardized all examples to 62 EEG channels. The paired target sentence was tokenized with the Whisper tokenizer using a maximum length of 64 tokens, and an input attention mask was stored to distinguish valid EEG samples from padded positions.
We considered two data-splitting protocols. In the \textbf{Random} split, sentence-level examples were randomly shuffled independently for each subject, and then divided into 80\% training, 10\% validation, and 10\% test examples. This setting evaluates within-subject generalization when training and test sets may contain different sentence segments from the same participant and may also contain the same sentence identities across different participants. To provide a more stringent evaluation, we further constructed a \textbf{Sentence} split. In this protocol, we first collected all unique sentence IDs across the dataset, randomly partitioned the sentence IDs into 80\%/10\%/10\% train/validation/test sets, and then assigned every subject-specific EEG sample to the split determined by its sentence ID. As a result, no sentence identity appears in more than one split, preventing the model from seeing neural responses to the same linguistic stimulus during training and testing.

\paragraph{MEG preprocessing and data splits.}
For the MEG-MASC dataset, we constructed sentence-level MEG-speech pairs from the original continuous recordings and event annotations. We parsed the BIDS-style events files and grouped word-level annotations according to their sequence identifiers, thereby obtaining sentence-level segments with the corresponding story metadata, word timings, speech boundaries, and MEG onset/offset times. For each recording, the raw KIT MEG file was loaded with MNE, and we retained only MEG channels while excluding miscellaneous, reference MEG, EEG, stimulus, EOG, and ECG channels. The MEG signal was notch-filtered at 50 Hz, band-pass filtered between 1 and 58 Hz, and resampled to 200 Hz. Each sentence-level MEG segment was then extracted according to the event timestamps. We further applied robust scaling using the initial portion of each segment as the reference, clipped values whose absolute magnitude exceeded 10, and normalized the clipped signal by this threshold. Segments with abnormal values, including empty arrays, NaNs, infinities, or excessive zeros, were excluded. The corresponding speech segment was extracted using the provided speech timestamps and resampled to 16 kHz.
We considered two splitting protocols. In the \textbf{Random} split, all sentence-level samples were randomly shuffled and divided into 80\% training, 10\% validation, and 10\% test sets. This split provides a standard random-sample evaluation but may contain the same sentence text across different splits because MEG-MASC includes repeated or shared linguistic stimuli. We therefore constructed the stricter \textbf{Sentence} split, where we normalized sentence text, grouped all samples by unique sentence identity across the entire dataset, and assigned each sentence group to exactly one of the train, valid, or test splits using an 80\%/10\%/10\% partition. Consequently, no sentence text appears in more than one split.

\subsection{Baselines}
\label{appendix:baselines}

\paragraph{Vanilla Baseline.}
We consider a vanilla brain-to-speech regression baseline that maps neural recordings to BigVGAN-compatible log-Mel spectrograms. The pipeline consists of three stages. First, for each sentence-level speech segment, we precompute the acoustic target using the standard BigVGAN Mel preprocessing procedure. The waveform is resampled to 22.05 kHz, transformed into a magnitude spectrogram with STFT, projected onto an 80-bin Mel filterbank, and compressed with a logarithmic dynamic-range transform. Second, we train a neural network to predict the Mel spectrogram from the corresponding EEG or MEG segment. To eliminate interference caused by model architecture, this neural network adopts almost the same design as the neural signal embedder and projector in our \textit{MindVoice}. Third, the model is optimized with a spectrogram MSE objective the same as \textit{MindVoice}. The same baseline method has been adopted in advanced research on speech neuroprosthetics~\cite{wairagkar2025instantaneous}, and the only difference is that they adopt a lightweight vocoder to achieve real-time reconstruction. 
For both EEG and MEG experiments, we train the vanilla baseline using the corresponding preprocessed train/validation splits. The model is optimized with AdamW and cosine learning-rate scheduling. Gradients are clipped to a maximum norm of 2.0, and the best checkpoint is selected according to validation loss with early stopping patience 8 epochs. For the Brennan EEG experiments, we train for at most 300 epochs with batch size 48, learning rate 1e-4, and weight decay 0.01. For the MEG-MASC experiments, we train for at most 100 epochs with batch size 32, learning rate 3e-4, and weight decay 0.01. In most training runs, training was terminated by early stopping.

\paragraph{FESDE baseline.}
We use FESDE~\citep{lee2024toward} as another speech reconstruction baseline. This framework consists of an neural signal module, a speech module, and a connector that bridges the latent spaces of neural signal and speech. The neural signal module follows an encoder-decoder design for neural representation learning: the encoder maps raw neural signals into latent embeddings using convolutional blocks and Structured State Space Sequence layers, while the decoder reconstructs the input neural signal through deconvolutional blocks. The speech module is based on VITS and contains a speech encoder and a HiFi-GAN speech decoder. The speech encoder takes linear spectrograms as input and produces the mean and variance of a speech latent distribution using WaveNet-style residual blocks, while the decoder generates the speech waveform from the latent speech embedding. The connector maps the learned neural embedding to the speech latent space through a Transformer-based prenet and a normalizing flow, providing an mapping between the neural embeddings and the speech latent. During training, FESDE combines an neural autoencode objective, a mel-spectrogram MSE loss between generated and ground-truth speech, a KL-divergence loss for aligning the neural embeddings with the speech latent, and HiFi-GAN adversarial loss. A gradient stop is applied between the neural embedding and the connector to stabilize optimization by decoupling neural representation learning from the early-stage speech-generation objective.
To adapt FESDE to our EEG and MEG datasets, we modified the model input interface while keeping the original FESDE training configuration whenever possible, including the loss weights, optimizer hyperparameters, and learning-rate schedule. We trained the model for 1,000 epochs on the EEG dataset and 300 epochs on the MEG dataset, and used early stopping with a patience of 8. Based on our reported results, the adapted FESDE baseline already achieves a better MCD (from 10.15 to 11.31) than the value reported in the original FESDE paper (from 11.43 to 11.46) on a different EEG dataset. Although the two results are not directly comparable due to differences in datasets and preprocessing, this suggests that our implementation is sufficiently competitive as a baseline method.

\subsection{Evalution Metrics}
\label{appendix:metrics}

\paragraph{Mel-MSE.}
Mel-MSE measures the frame-level discrepancy between the mel spectrogram of the reconstructed waveform and the ground-truth waveform. In our implementation, both waveforms are first converted into Whisper-style log-Mel input features using the Whisper processor at a sampling rate of 16 kHz. To ensure that the comparison is made over the reference duration, we truncate both predicted and reference Mel features to the number of frames determined by the ground-truth waveform length. The final score is averaged over all test utterances. Lower Mel-MSE indicates that the reconstructed waveform is closer to the reference waveform in the log-Mel acoustic feature space.

\paragraph{MCD.}
Mel-cepstral distortion (MCD) measures the spectral-envelope distance between reconstructed and ground-truth speech in the mel-cepstral domain. We extract mel-cepstral coefficient sequences from both waveforms, align the two sequences with dynamic time warping, and compute the average frame-wise Euclidean distance after excluding the energy coefficient $c_0$. For an aligned sequence of length $T$, MCD is computed as
\[
\mathrm{MCD}
=
\frac{10\sqrt{2}}{\ln 10}
\cdot
\frac{1}{T}
\sum_{t=1}^{T}
\sqrt{
\sum_{d=1}^{D}
\left(c_{t,d}-\hat{c}_{t,d}\right)^2
}.
\]
Here, $c_{t,d}$ and $\hat{c}_{t,d}$ denote the reference and reconstructed mel-cepstral coefficients, respectively. The final MCD is averaged over all test utterances. Lower MCD indicates smaller spectral distortion.

\paragraph{HuBert.}
HuBert measures whether the reconstructed and reference waveforms have similar semantic-level speech representations. HuBERT is a self-supervised speech representation model trained by predicting hidden-unit cluster assignments over masked speech regions, which encourages the model to encode both semantic and acoustic structure in continuous speech. In our implementation, we use \texttt{hubert-base-ls960} model. Each waveform is passed through the corresponding feature extractor at 16 kHz and then through the HuBERT model. We take the last hidden states and average them over the temporal dimension to obtain one embedding vector. The HuBert score is the cosine similarity between predicted and reference embeddings. The final metric is averaged over all utterances. Higher HuBert metirc indicates that the reconstructed waveform better preserves the speech content of the ground-truth speech.

\paragraph{Baseline-scaled ASR-BERTScore-F1.}
This metric evaluates intelligibility at the text level by first transcribing the reconstructed speech and then comparing the resulting transcript with the ground-truth transcript using BERTScore. In our implementation, reconstructed speech files are transcribed with \texttt{Qwen3-ASR-0.6B}. We then compute BERTScore-F1 between the reconstructed and ground-truth transcripts using \texttt{roberta-large} with outputs from its 17th layer. BERTScore matches contextual token embeddings between candidate and reference sentences using cosine similarity, and aggregates these matches into precision, recall, and F1 scores; we report F1 as ASR-BERTScore-F1. Since raw BERTScore values are often concentrated in a narrow high-value range, especially for RoBERTa-based models, we apply baseline rescaling for clearer comparison. Specifically, we subtract the random-sentence baseline provided by the BERTScore GitHub repository and normalize the score so that the baseline maps to 0 and a perfect match maps to 1. As a result, individual ASR-BERTScore values can be negative when the reconstructed transcript is less similar to the reference than the random baseline. For empty ASR outputs, we directly assign the corresponding baseline value before rescaling. The final score is the average scaled F1 across all utterances. A higher score indicates that the reconstructed speech is more intelligible and semantically closer to the ground-truth transcription.

\paragraph{WavLM.}
WavLM measures whether the reconstructed and ground truth waveforms preserve similar speaker-related and timbre-level speech representations. WavLM is a self-supervised speech representation model designed for full-stack speech processing, with an emphasis on both spoken-content modeling and speaker identity preservation. In particular, WavLM has been used for speaker verification, making it suitable for evaluating whether reconstructed speech preserves voice characteristics such as timbre. In our implementation, we use the \texttt{wavlm-base-sv} model, which is specifically adapted for speaker verification. Each waveform is passed through the corresponding feature extractor at 16 kHz and then through the WavLM model. We take the last hidden states and average them over the temporal dimension to obtain one embedding vector. The WavLM score is the cosine similarity between predicted and reference embeddings. The final metric is averaged over all utterances. A higher WavLM metric indicates that the reconstructed waveform better preserves the speaker identity and timbre characteristics of the ground-truth speech.

\paragraph{MOS.}
We report MOS using an automatic MOS predictor model rather than human listening tests. Specifically, we use the \texttt{utmos22\_strong} model from SpeechMOS/UTMOS, which was developed for automatic prediction of human mean opinion scores in the VoiceMOS Challenge 2022. UTMOS predicts the perceptual quality of a speech waveform, where a higher score indicates better naturalness and overall speech quality. In our implementation, we feed each reconstructed waveform into the MOS predictor at a sampling rate of 16 kHz and report their average value.

\subsection{Implementation Details}
\label{appendix:implementation}

The speech VQ-VE uses a convolutional encoder-decoder architecture with residual blocks, group normalization, swish activations, and self-attention layers. The speech vocabulary contains $2048$ entries. We use a $4\times$ downsampling module for input mel-spectrogram. The neural embedder maps neural recordings to this latent space using a temporal-spatial convolutional frontend followed by learnable positional embeddings and a Transformer encoder. The Transformer encoder has $4$ layers with $256$ hidden size, $8$ attention heads, and $2048$ feed-forward dimension. All models were implemented in PyTorch. We optimized all trainable parameters using AdamW with a weight decay of $0.01$, and adopted a cosine annealing learning-rate scheduler for all training runs. Model selection was based on the validation loss, and early stopping was applied with a patience of $8$. On the MEG dataset, the model was trained with a batch size of $16$ and a learning rate of 3e-4 for up to $50$ epochs; on the EEG dataset, we used the same batch size and learning rate, and trained the model for up to $100$ epochs. The loss weights $\alpha_1$ and $\alpha_2$ were both set to $0.5$. To improve robustness to neural inputs, we additionally included silence samples as data augmentation. For the neuro-to-semantic aligner, in the first stage, we used a batch size of $8$ and a learning rate of 1e-4 on the MEG dataset, and a batch size of $16$ with the same learning rate on the EEG dataset. The AdamW momentum parameters $\beta_1$ and $\beta_2$ were both set to $0.1$. The same hyperparameter settings were used in the second stage, and both stages were trained for $300$ epochs. For acoustic-level model training, this model was trained with a batch size of $32$, a learning rate of 1e-4, and $50$ epochs, with the loss weight $\gamma$ set to $0.1$. During text reconstruction, we used beam search with a beam size of $5$, a maximum sequence length of $128$, a maximum of $30$ newly generated tokens, and a minimum of $1$ newly generated token. We further applied a repetition penalty of $5.0$, a no-repeat $2$-gram constraint, and a length penalty of $1.2$. For speech generation, we generated one speech sample per test utterance using top-$p=0.5$, temperature $0.5$, repetition penalty $1.1$, and chunk length $512$. The generated waveform was resampled to $16\,\mathrm{kHz}$ before evaluation. All experiments were conducted on NVIDIA Tesla V100 GPUs. A single training run took approximately $20$ hours on the EEG dataset and approximately $40$ hours on the MEG dataset.

\begin{figure}[!t]
    \centering
    \includegraphics[width=0.98\linewidth]{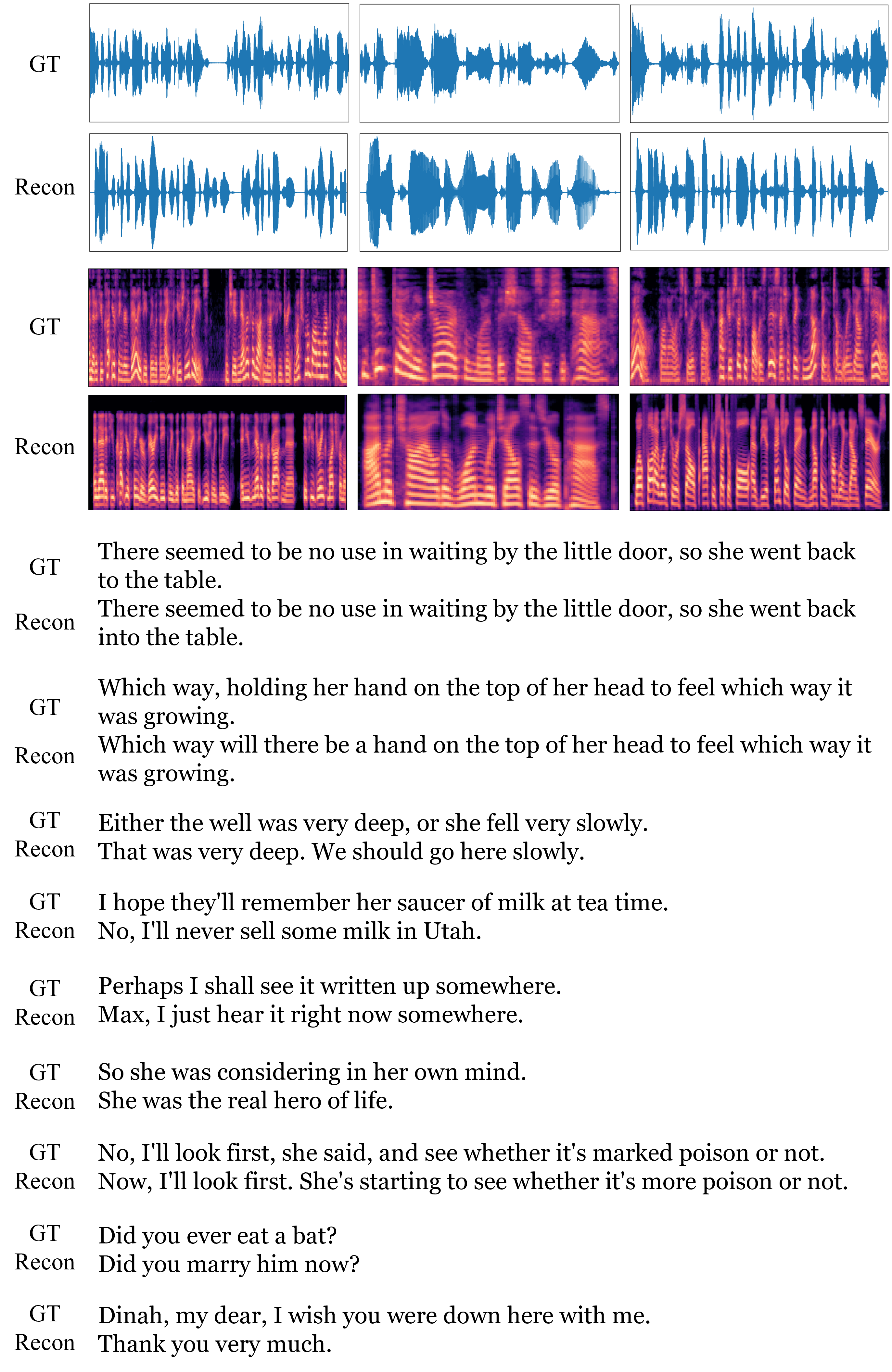}
    \vspace{-1mm}
    \caption{Additional qualitative results on the Brennan EEG dataset. Randomly select.}
    \label{more_eeg_result}
\end{figure}

\begin{figure}[!t]
    \centering
    \includegraphics[width=0.99\linewidth]{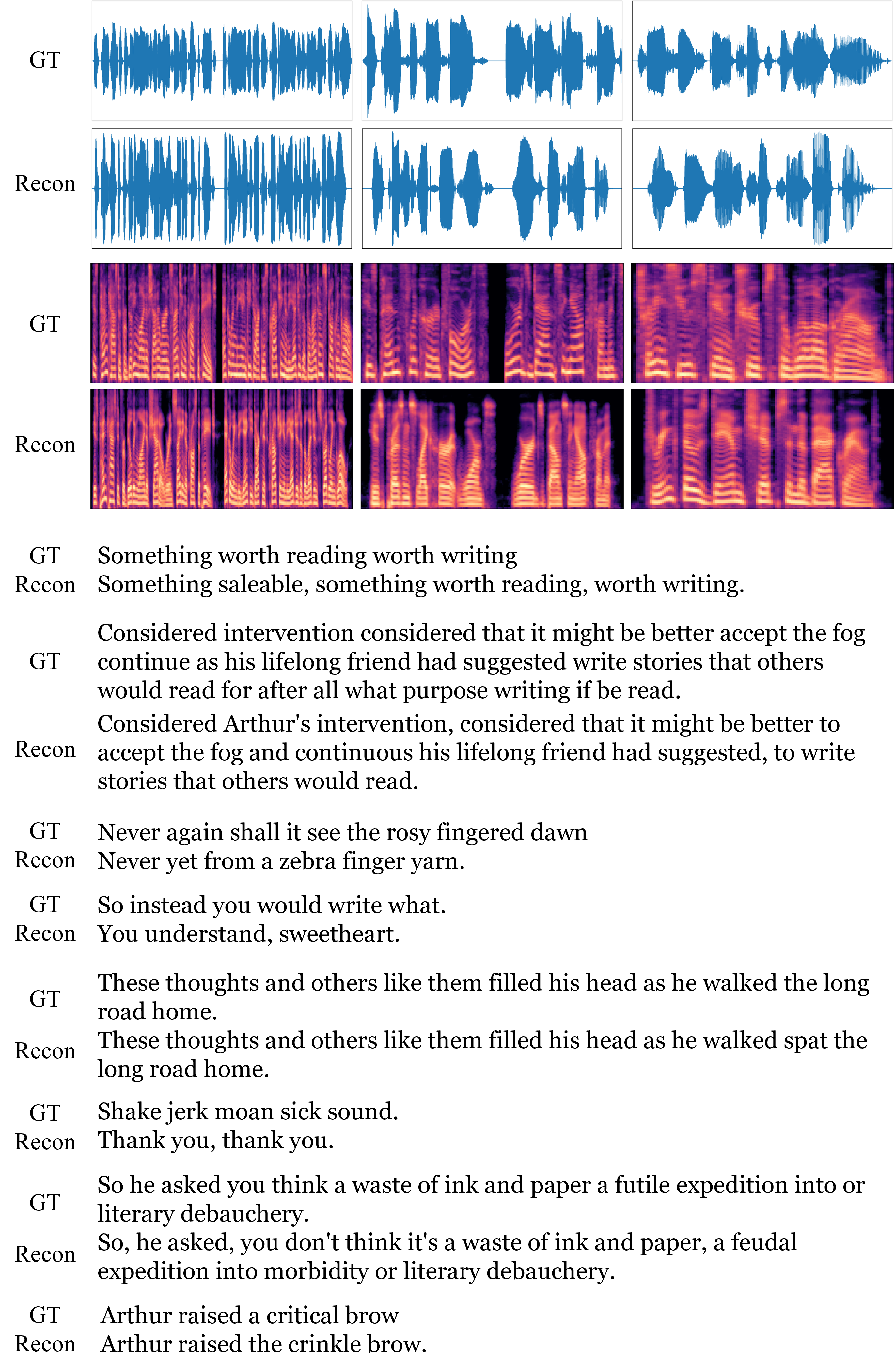}
    \vspace{-1mm}
    \caption{Additional qualitative results on the Gwilliams MEG dataset. Randomly select.}
    \label{more_meg_result}
\end{figure}

\section{Additional Results}
\subsection{Additional Qualitative Results}
\label{appendix:qualitative}
We present additional qualitative visualization results in Figure~\ref{more_eeg_result} and~\ref{more_meg_result}, corresponding to the EEG and MEG datasets, respectively. It can be observed that the speech waveforms reconstructed by our method exhibit a certain degree of similarity to the ground-truth waveforms in terms of amplitude variations and rhythmic patterns. The addtional transcription results show that our method can recover meaningful semantic content to some extent. In some cases, the reconstructed semantics are highly consistent with the ground truth, indicating that the model is capable of producing intelligible speech with correct or closely related meanings. However, in many cases, the reconstructed outputs tend to preserve only a few salient keywords rather than the complete sentence-level semantics. In more challenging cases, the reconstruction may even degenerate into fixed phrases with completely incorrect meanings, reflecting the instability of the current model under certain neural signal conditions. Overall, these results suggest that reconstructing intelligible continuous speech from neural signals remains a challenging task, and further improvements are still needed.

\begin{figure}[!t]
    \centering
    \includegraphics[width=1.0\linewidth]{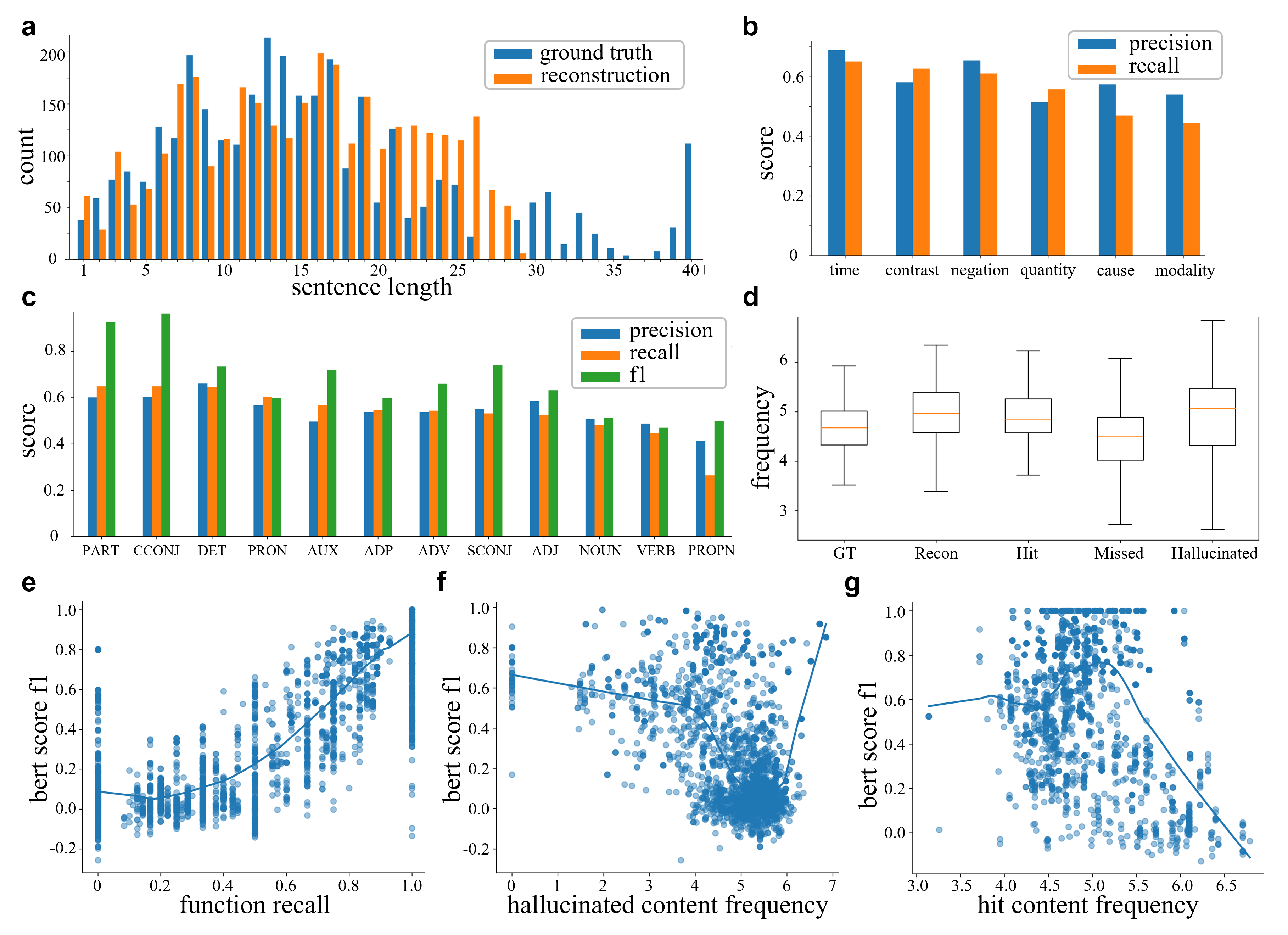}
    \vspace{-3mm}
    \caption{\textbf{Fine-grained linguistic preference analysis of MEG-to-text reconstruction}. \textbf{a}, ground-truth and reconstructed sentence-length distributions. \textbf{b}, precision and recall of discourse/relation markers. \textbf{c}, POS-specific precision, recall, and F1. \textbf{d}, zipf frequency distributions of content words in five groups: GT, Recon, Hit, Missed, and Hallucinated. \textbf{e–g}, associations between BERTScore and function-word recall, hallucinated-word frequency, and hit-word frequency.}
    \label{more_prefer}
    \vspace{-3mm}
\end{figure}

\subsection{Additional Perference Analyses}
\label{appendix:perference}

To characterize the linguistic preferences of the MEG-to-text decoder beyond aggregate BERTScore, we conduct a fine-grained analysis of reconstructed sentences. Figure~\ref{more_prefer}{\color{blue} a} compares the sentence-length distributions of ground-truth and reconstructed sentences. The two distributions largely overlap, indicating that the decoder captures coarse sentence-length information rather than producing arbitrary-length outputs. However, the reconstructed distribution is more concentrated around medium lengths and contains fewer long sentences, suggesting a tendency to under-reconstruct long targets and regress toward typical sentence lengths.
We next examine which linguistic units are more faithfully recovered. Figure~\ref{more_prefer}{\color{blue} b} evaluates the reconstruction of discourse and relation markers. Temporal, contrastive, and negation markers are partially preserved, whereas causal and modal markers show lower recall. This suggests that the decoder can recover some explicit relation cues, but more abstract relational information and speaker-attitude information remain more fragile. Since this analysis is based on lexical marker overlap, it measures preservation of relation cues rather than full compositional correctness. Figure~\ref{more_prefer}{\color{blue} c} reports POS-specific precision, recall, and F1 scores, where words are lemmatized and grouped by spaCy POS tags. The results show that reconstruction fidelity varies substantially across POS categories: grammatical function categories are generally recovered more reliably than lexically specific content categories. This indicates that the decoder captures the grammatical scaffold of the sentence more robustly than exact lexical identity.

We further analyze lexical frequency bias in content-word reconstruction. In our implementation, content words are defined as nouns, proper nouns, verbs, adjectives, and adverbs. Let $G$ be the set of content words in the ground-truth sentence and $R$ be the set of content words in the reconstruction. We define $ \mathrm{Hit} = G \cap R$, $\mathrm{Missed} = G \setminus R$, and $\mathrm{Hallucinated} = R \setminus G$. Figure~\ref{more_prefer}{\color{blue} d} compares the Zipf frequencies of content words across these groups. Recovered words are more frequent than missed words, and hallucinated words are also biased toward high-frequency vocabulary. This reveals a clear lexical frequency bias: high-frequency content words are more likely to be recovered, while low-frequency content words are more likely to be omitted. Moreover, when the decoder introduces additional content words, they tend to be frequent and generic rather than rare or specific.
Finally, Figure~\ref{more_prefer}{\color{blue} e}-\ref{more_prefer}{\color{blue} g} relate these fine-grained properties to BERTScore. Figure~\ref{more_prefer}{\color{blue} e} shows a strong positive association between function-word recall and BERTScore, indicating that high-quality reconstructions preserve not only semantic content but also the grammatical scaffolding that organizes semantic relations. Figure~\ref{more_prefer}{\color{blue} f} shows that low-scoring reconstructions often contain high-frequency hallucinated content words, consistent with the decoder falling back to generic lexical choices when the target content is uncertain. Figure~\ref{more_prefer}{\color{blue} g} shows a non-monotonic relationship between the frequency of recovered content words and BERTScore: recovering moderately frequent content words improves reconstruction quality, whereas relying primarily on extremely frequent words may reflect generic semantic reconstruction rather than faithful lexical recovery. Together, these analyses suggest that the decoder preserves coarse length, syntactic, and discourse structure, but exact lexical reconstruction is biased toward frequent and generic words.

\end{document}